\documentclass[twocolumn,aps,superscriptaddress]{revtex4}

\usepackage{amssymb}
\usepackage{amsmath}
\usepackage{graphicx}
\usepackage[normalem]{ulem}
\usepackage[dvips]{color}
\usepackage{appendix}

\begin{document}

\title{Nucleon spin polarization in intermediate-energy heavy-ion collisions}
\author{Yin Xia}
\affiliation{Institute of Applied Physics and Computational Mathematics, Beijing 100094, China}
\author{Jun Xu\footnote{corresponding author: xujun@zjlab.org.cn}}
\affiliation{Shanghai Advanced Research Institute, Chinese Academy of Sciences,
Shanghai 201210, China}
\affiliation{Shanghai Institute of Applied Physics, Chinese Academy
of Sciences, Shanghai 201800, China}

\date{\today}

\begin{abstract}
Based on a spin-dependent Boltzmann-Uehling-Uhlenbeck transport model with spin-dependent potentials incorporated using the lattice Hamiltonian method, we have studied the global spin polarization perpendicular to the reaction plane as well as the local spin polarization in the longitudinal direction in non-central intermediate-energy heavy-ion collisions, as an extension of similar studies in relativistic heavy-ion collisions. Both the global and the local spin polarizations are found to be mostly dominated by the time-odd component of the nuclear spin-orbit potential. Impacts of various theoretical uncertainties on the nucleon spin polarization as well as its dependence on the beam energy and impact parameter are discussed. Our study serves as a baseline for understanding the spin polarization mechanism of strongly interacting matter produced in heavy-ion collisions dominated by nucleon degree of freedom.
\end{abstract}

\maketitle

The spin physics is an interesting topic in various fields. Due to the spin-orbit coupling, the spin polarization phenomena can generally be induced in a vorticity field. In non-central relativistic heavy-ion collisions, it was pointed out~\cite{Lia05} that the produced quark-gluon plasma can be globally polarized perpendicular to the reaction plane, as a result of the angular momentum transfer from colliding nuclei to the midrapidity region. This can lead to the spin polarization of $\Lambda$ hyperons or vector mesons, whose polarizations can be experimentally measurable through the angular distribution of their decays. Although the experimentally measured spin polarizations of these particles are small at ultrarelativistic energies~\cite{STAR07,STAR08,STAR18}, they become considerably larger at lower collision energies~\cite{STAR17}. On the theoretical side, it was found that although the spin polarization is affected by various properties of the heavy-ion system~\cite{Bet07}, it can be well described by the local  thermodynamic equilibrium condition and hydrodynamics~\cite{Bec08a,Bec08b}. Analytically, the polarizations of hadrons are generally expressed as those on the Cooper-Frye freeze-out hypersurface~\cite{Bec13a,Bec13b,Fan16}. Using transport simulations, it was found that the strength of the vorticity field decreases with the increasing collision energy~\cite{Jia16}, consistent with the experimentally observed stronger spin polarizations at lower collision energies~\cite{STAR17}. There are further follow-up theoretical efforts~\cite{Li17,Sun17,Xie17,Bec17} based on transport models or hydrodynamic models devoted to give quantitative explanations to the spin polarization phenomena, showing that the quark-gluon plasma can be the most vortical fluid on the earth. Besides the global spin polarization, theorists have more recently predicted that there could be some local structures of the vorticity field~\cite{Pan16}, and they can lead to the azimuthal angular dependence of the longitudinal spin polarization~\cite{Xia18,Sun19,Bec18}. However, the latest experimental data from the STAR Collaboration shows a different azimuthal angular dependence~\cite{STAR19} compared to most theoretical studies. Very recently, the feed-down effect in the hadronic afterburner on the $\Lambda$ spin polarization has also drawn considerable attentions~\cite{Bec17,Xia19,Bec19}. These spin polarization phenomena in heavy-ion collisions serve as good probes of understanding the vortical dynamics in the strongly interacting matter as well as the properties of the spin-orbit coupling.

In intermediate-energy heavy-ion collisions, one expects that the spin polarization phenomena should exist as well, as a result of the nuclear spin-orbit interaction, which was first proposed to explain the magnetic number of finite nuclei~\cite{May48,Hax49}. Except for the well-known importance of the nuclear spin-orbit coupling in nuclear structure studies, it also has dramatic dynamic effects in low- and intermediate-energy heavy-ion collisions  (see, e.g., Ref.~\cite{Xu15} for a review). For instance, the nuclear spin-orbit interaction may enhance dissipations and can affect significantly the fusion threshold in low-energy heavy-ion reactions based on studies using the time-dependent Hartree-Fock model~\cite{Uma86,Dai14}. Using a spin-dependent Boltzmann-Uehling-Uhlenbeck (SBUU) transport model, it was found that the nuclear spin-orbit coupling leads to different dynamics and thus different collective flows of spin-up and spin-down nucleons~\cite{Xu13,Xia14}, with respective to the direction perpendicular to the reaction plane. With an improved SBUU model based on the lattice Hamiltonian framework, we are going to study the global and local nucleon spin polarization induced by the nuclear spin-orbit coupling in intermediate-energy heavy-ion collisions.

In the SBUU approach, the initial nucleon coordinates are sampled according to the radial distribution generated by the Skyrme-Hartree-Fock calculation~\cite{Vau72,Che10}, while their momenta are sampled isotropically within the Fermi momentum according to the local density. The coordinates and momenta are then boosted in the longitudinal ($z$) direction according to the beam energy. We use the convention that the $xoz$ plane is the reaction plane, and the $xoy$ plane is the transverse plane. The expectation spin vector $\vec{\sigma}$ of each nucleon is represented as a unit vector, from which the probability of spin-up and spin-down states with respective to an arbitrary direction can be obtained, and their initial directions are randomized in the $4\pi$ direction. With the initial condition set up, the time evolution of each nucleon is determined by the mean-field potential as well as the nucleon-nucleon collisions described in the following.

The spin-dependent mean-field potential originates from the following Skyrme-type effective spin-orbit interaction between two nucleons at positions $\vec{r}_1$ and $\vec{r}_2$~\cite{Vau72}
\begin{equation}
v_{so} = i W_0 (\vec{\sigma}_1+\vec{\sigma}_2) \cdot \vec{k}^\prime \times
\delta(\vec{r}_1-\vec{r}_2) \vec{k},
\end{equation}
where $W_0$ is the strength of the spin-orbit coupling whose default value is set to be 150 MeVfm$^5$ in this study, $\vec{\sigma}_{1(2)}$ is the Pauli matrix, $\vec{k}=(\vec{p}_1-\vec{p}_2)/2$ is the relative momentum operator acting on the right with $\vec{p}=-i\nabla$, and $\vec{k}^\prime$ is the complex conjugate of $\vec{k}$. With the Hartree-Fock method, the above spin-orbit interaction leads to the potential energy density functional expressed as~\cite{Eng75}
\begin{eqnarray}\label{vso}
V_{so} &=& V_{so}^0 + \sum_\tau V_{so}^\tau \notag\\
&=&-\frac{W_0}{2}[\rho \nabla \cdot \vec{J} + \vec{s} \cdot \nabla \times \vec{j} \notag\\
&+& \sum_\tau (\rho_\tau \nabla \cdot \vec{J}_\tau + \vec{s}_\tau \cdot \nabla \times \vec{j}_\tau)],
\end{eqnarray}
where $\tau=n,p$ is the isospin index, and
\begin{eqnarray}
\rho &=& \sum_\tau \rho_\tau = \sum_\tau \sum_i \phi^\star_i \phi_i,\\
\vec{s} &=& \sum_\tau \vec{s}_\tau = \sum_\tau\sum_i \sum_{\sigma,\sigma^\prime} \phi^\star_i
\langle\sigma|\vec{\sigma}|\sigma^\prime\rangle \phi_i, \\
\vec{j} &=& \sum_\tau \vec{j}_\tau = \frac{1}{2i} \sum_\tau\sum_i (\phi^\star_i \nabla\phi_i- \phi_i
\nabla\phi^\star_i),\\
\vec{J} &=& \sum_\tau \vec{J}_\tau = \frac{1}{2i} \sum_\tau\sum_i \sum_{\sigma,\sigma^\prime}
(\phi^\star_i \nabla\phi_i- \phi_i \nabla\phi^\star_i)\times
\langle\sigma|\vec{\sigma}|\sigma^\prime\rangle,\notag\\
\end{eqnarray}
are respectively the number, spin, momentum, and spin-current densities, with $\phi_i$ being the wave function of the $i$th nucleon. Those relevant to $\rho$ and $\vec{J}$ are the time-even terms, while those relevant to $\vec{s}$ and $\vec{j}$ are the time-odd terms~\cite{Eng75}. These densities can be expressed in terms of the spin-dependent phase-space distribution functions (see Eqs.~(14-17) in Ref.~\cite{Xia16}). Using the test-particle method~\cite{Won82,Ber88}, the spin-dependent phase-space distribution functions can be calculated from averaging over parallel events, with the contribution of each test particle a $\delta$ function in coordinate and momentum space (see Eqs.~(44-49) in Ref.~\cite{Xia16}). In the present study, we make an improvement by using the lattice Hamiltonian method~\cite{Len89} to calculate these densities. The average number, spin, momentum, and spin-current densities at the sites of a three-dimensional cubic lattice are defined respectively as
\begin{eqnarray}
\rho_L(\vec{r}_{\alpha})&=&\sum_{i}S(\vec{r}_{\alpha}-\vec{r}_i),\\
\vec{s}_L(\vec{r}_{\alpha})&=&\sum_{i}\vec{\sigma}_iS(\vec{r}_{\alpha}-\vec{r}_i),\\
\vec{j}_L(\vec{r}_{\alpha})&=&\sum_{i}\vec{p}_iS(\vec{r}_{\alpha}-\vec{r}_i),\\
\vec{J}_L(\vec{r}_{\alpha})&=&\sum_{i}\left(\vec{p}_i \times \vec{\sigma}_i\right)S(\vec{r}_{\alpha}-\vec{r}_i).
\end{eqnarray}
In the above, $\vec{p}_i\sim-i\nabla$ and $\vec{\sigma}_i\sim\langle\sigma|\vec{\sigma}|\sigma^\prime\rangle$ are the momentum and the spin expectation direction of the $i$th test particle, $\alpha$ is a site index and $\vec{r}_{\alpha}$ is the position of site $\alpha$, and $S$ is the shape function describing the contribution of a test particle at $\vec{r}_i$ to the average density at $\vec{r}_{\alpha}$, i.e.,
\begin{eqnarray}
S(\vec{r})=\frac{1}{N(nl)^6}g(x)g(y)g(z)
\end{eqnarray}
with
\begin{eqnarray}
g(q)=(nl-|q|)\Theta(nl-|q|).
\end{eqnarray}
$N$ is the number of parallel events, $l$ is the lattice spacing, $n$ determines the range of $S$, and $\Theta$ is the Heaviside function. We adopt $N=200$, $l=1$ fm, and $n=2$ in the present study. The Hamiltonian of the whole system is thus
\begin{equation}
H=\sum_{i}\frac{\vec{p}_{i}^{2}}{2m}+NV,
\end{equation}
where $m$ is the nucleon mass, and
\begin{equation}
NV=Nl^3\sum_\alpha [V_{MID}(\vec{r}_{\alpha}) + V_{so}(\vec{r}_{\alpha})]
\end{equation}
is the total potential energy, with $V_{MID}$ being a Skyrme-type potential energy corresponding to a momentum-independent mean-field potential, which depends only on the nucleon number density and is fitted to reproduce the empirical nuclear matter properties. The time evolution of the coordinate, momentum, and the spin of the $i$th nucleon with isospin $\tau$ is determined by the canonical equations of motion written as
\begin{eqnarray}
\frac{d\vec{r}_i}{dt} &=& \frac{\partial H}{\partial \vec{p}_i} = \frac{\vec{p}_i}{m} + Nl^3 \sum_\alpha \frac{\partial V_{so}}{\partial \vec{p}_i} \notag\\
&=& \frac{\vec{p}_i}{m} -  Nl^3 \frac{W_0}{2} \sum_\alpha \{\rho(\vec{r}_{\alpha}) [\vec{\sigma}_i \times \nabla S(\vec{r}_{\alpha}-\vec{r}_i)]  \notag\\
&+& \vec{s}(\vec{r}_{\alpha})  \times \nabla S(\vec{r}_{\alpha}-\vec{r}_i) \} + Nl^3 \sum_\alpha \frac{\partial V_{so}^\tau}{\partial \vec{p}_i}, \\
\frac{d\vec{p}_i}{dt} &=& -\frac{\partial H}{\partial \vec{r}_i} = - Nl^3 \sum_\alpha\frac{\partial V_{MID}}{\partial \vec{r}_i} - Nl^3 \sum_\alpha\frac{\partial V_{so}}{\partial \vec{r}_i} \notag\\
&=& - Nl^3 \sum_\alpha\frac{\partial V_{MID}}{\partial \vec{r}_i} \notag\\
&+& Nl^3 \frac{W_0}{2} \sum_\alpha \{ \frac{\partial S(\vec{r}_{\alpha}-\vec{r}_i)}{\partial \vec{r}_i} \nabla \cdot \vec{J}(\vec{r}_{\alpha}) \notag\\
&+& \rho(\vec{r}_{\alpha}) \nabla \frac{\partial S(\vec{r}_{\alpha}-\vec{r}_i)}{\partial \vec{r}_i} \cdot (\vec{p}_i \times \vec{\sigma}_i) \notag\\
&+&  \frac{\partial S(\vec{r}_{\alpha}-\vec{r}_i)}{\partial \vec{r}_i} \vec{\sigma}_i \cdot [\nabla \times \vec{j}(\vec{r}_{\alpha})] \notag\\
&+& \vec{s}(\vec{r}_{\alpha}) \cdot [\nabla\frac{\partial S(\vec{r}_{\alpha}-\vec{r}_i)}{\partial \vec{r}_i} \times \vec{p}_i] \} - Nl^3 \sum_\alpha\frac{\partial V_{so}^\tau}{\partial \vec{r}_i},\\
\frac{d\vec{\sigma}_i}{dt} &=& \frac{1}{i}[\vec{\sigma}_i, H] \notag\\
&=& -Nl^3 W_0 \sum_\alpha \{ \rho(\vec{r}_{\alpha}) [\nabla S(\vec{r}_{\alpha}-\vec{r}_i) \times \vec{p}_i] \times \vec{\sigma}_i \notag\\
&+& [\nabla \times \vec{j}(\vec{r}_{\alpha})]\times \vec{\sigma}_iS(\vec{r}_{\alpha}-\vec{r}_i)\} +\frac{1}{i}[\vec{\sigma}_i, Nl^3\sum_\alpha V_{so}^\tau ], \notag\\\label{proc}
\end{eqnarray}
where $\nabla$ means taking the spatial derivative with respect to $\vec{r}_\alpha$, and the $V_{so}^\tau$ terms are the contributions of the third and the fourth terms in Eq.~\eqref{vso} depending on the nucleon isospin $\tau$ with the same structure.

Besides the soft mean-field potential, the dynamics is also affected by hard two-body nucleon-nucleon collisions basically following Bertsch's prescription~\cite{Ber88}. We employ the spin-dependent differential cross sections for the spin-singlet and spin-triplet states extracted from the phase-shift analyses of nucleon-nucleon scattering data~\cite{Arn77}. The spin state is determined with respective to the angular momentum of the two colliding nucleons, and the energy and angular dependence of the cross sections are parameterized in Ref.~\cite{Xia17}.
Whether the nucleon-nucleon collision is successful is determined by a spin- and isospin-dependent Pauli blocking, with the local phase-space cell more specific for nucleons with different spin and isospin states.

Since the nucleon spin polarization can be affected not only by the spin-dependent mean-field potential but also by spin-dependent collisions, we use two scenarios to determine the nucleon spin after a nucleon-nucleon collision. In one scenario, the nucleon spin is unchanged after collisions, representing the extreme case with the pure mean-field potential effect. In the more realistic case, nucleon-nucleon collisions will affect the nucleon spin evolution, but how the nucleon spin is changed after collisions is largely
unknown, especially in nuclear medium. To effectively take the above effect into consideration, we also use another scenario, where the spin of the nucleon with isospin $\tau$ rotates around the vector
\begin{eqnarray}
\vec{I}_i &=& -Nl^3 W_0 \sum_\alpha \{ [\rho(\vec{r}_{\alpha})+\rho_\tau(\vec{r}_{\alpha})] [\nabla S(\vec{r}_{\alpha}-\vec{r}_i) \times \vec{p}_i] \notag\\
&+& [\nabla \times \vec{j}(\vec{r}_{\alpha})+\nabla \times \vec{j}_\tau(\vec{r}_{\alpha})]S(\vec{r}_{\alpha}-\vec{r}_i)\}
\end{eqnarray}
for a random angle after collisions, and meanwhile satisfies the energy conservation condition according to the equation [Eq.~\eqref{proc}], which can be simply written as $d\vec{\sigma}_i/dt = \vec{I}_i \times \vec{\sigma}_i$ describing the processional motion of the nucleon spin. We will see how different collision treatments representing uncertainties of the in-medium nucleon spin flip after nucleon-nucleon collisions affect the nucleon spin polarization in intermediate-energy heavy-ion collisions.

\begin{figure*}[ht]
	\includegraphics[scale=0.3]{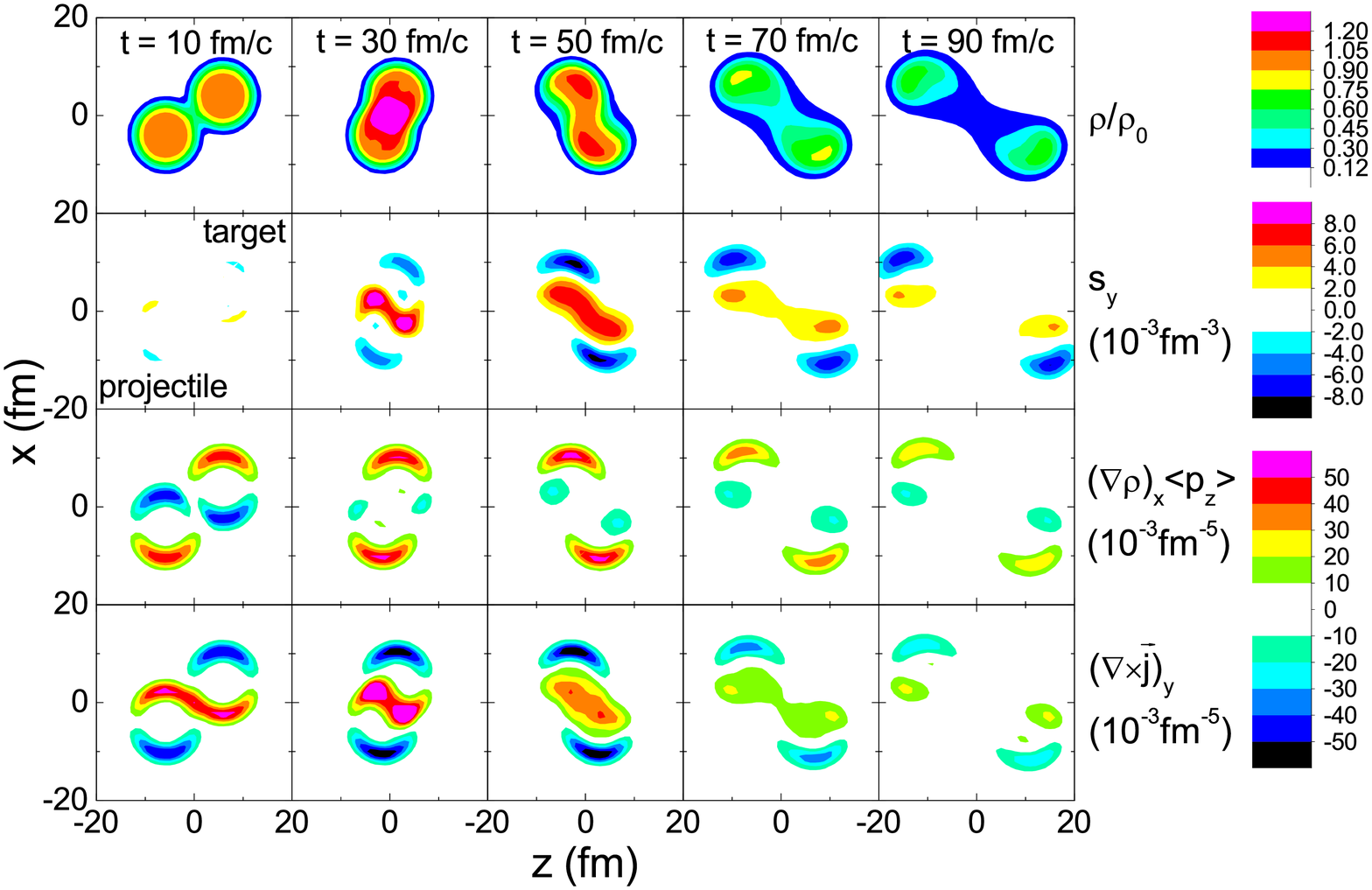}\includegraphics[scale=0.3]{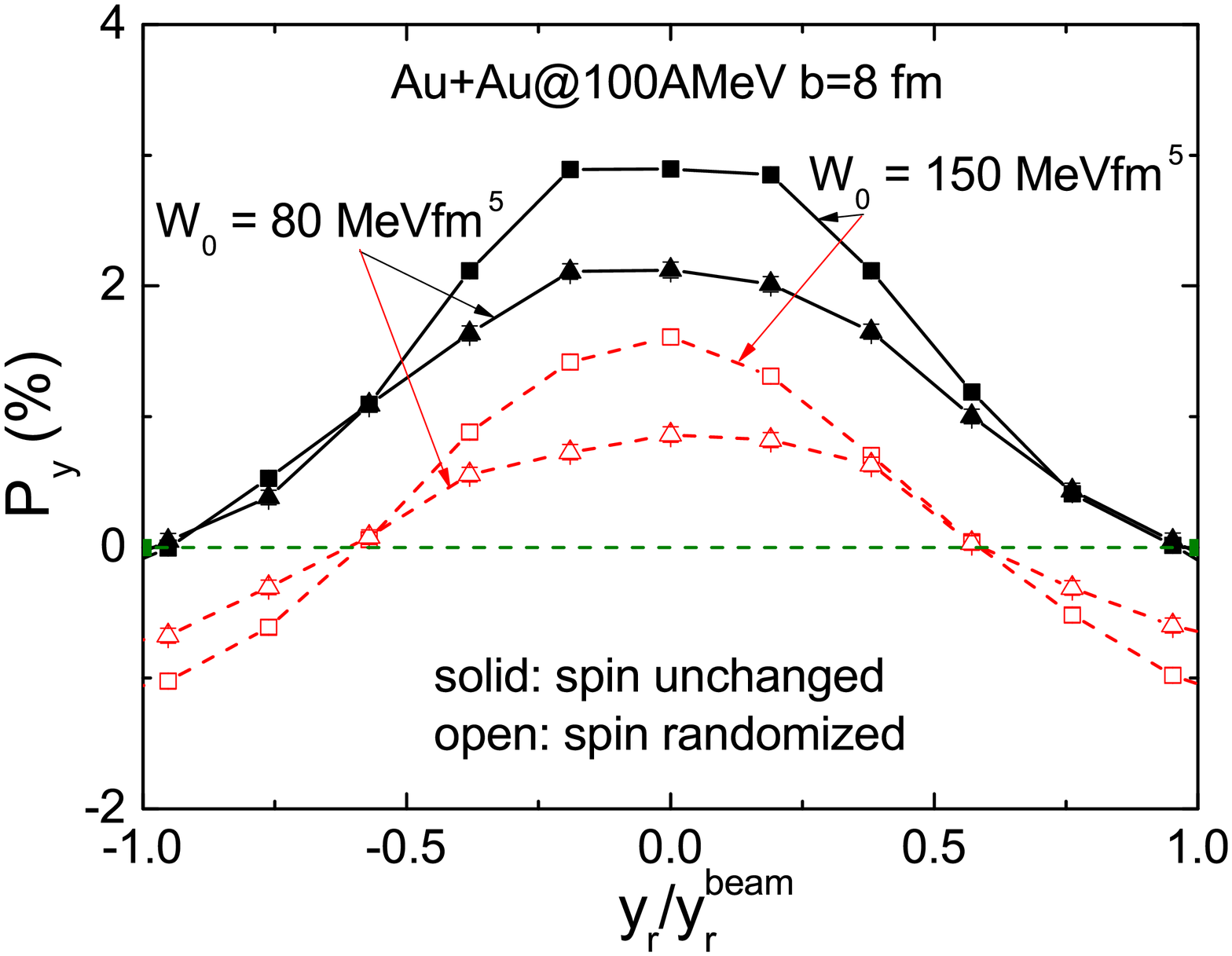}
	\caption{(Color online) Left: Density contours of the reduced nucleon number density (first row), $y$ component of the nucleon spin density (second row), $y$ component of the time-even spin-orbit potential (third row), and $y$ component of the time-odd spin-orbit potential in the reaction plane, for the spin unchanged scenario; Right: Spin polarization in $y$ direction of free nucleons with different strengths of the spin-orbit potential and collision treatments as a function of the reduced rapidity. Results are for Au+Au collisions at the beam energy of 100 AMeV and an impact parameter $b=8$ fm.} \label{xoz}
\end{figure*}

We first investigate the nucleon spin polarization in a typical system, i.e., Au+Au collisions at the beam energy of 100 AMeV and an impact parameter $b=8$ fm. The time evolutions of the reduced nucleon number density and the $y$ component of the nucleon spin density in the reaction plane are shown in the first and the second row in the left panel of Fig.~\ref{xoz}, respectively. It is seen that the participant matter is polarized in the $+y$ direction along the angular momentum of the collision system, while the spectator matter is polarized in the $-y$ direction antiparallel to the angular momentum. To understand the mechanism of this spin polarization, we further plot the $y$ components of the time-even and the time-odd spin-orbit potential in the third and the fourth row in the left panel of Fig.~\ref{xoz}, respectively, with the former dominated by the $(\nabla \rho)_x \langle p_z \rangle$ contribution. It is seen that the time-odd potential has the opposite but larger effect compared to the time-even potential, and generates the spin polarization shown in the second row. Similar mechanism was found when we studied the spin-dependent collective flows in intermediate-energy heavy-ion collisions~\cite{Xu13,Xia14}. These density contours are for the scenario with the nucleon spin unchanged after collisions. If the nucleon spin is randomized after collisions, the density contours are similar except that the spin polarization of the participant matter is partially washed out and becomes weaker. The rapidity distributions of the global spin polarization in $y$ direction of free nucleons, which is defined as those with a local density lower than 1/8 the saturation density $\rho_0=0.16$ fm$^{-3}$, are displayed in the right panel of Fig.~\ref{xoz} for different scenarios. It is seen that the spin polarization is generally stronger at midrapidities dominated by nucleons mostly emitted from the participant matter, but weaker at larger rapidities attributed to nucleons emitted from both participant and spectator matters, consistent with that observed in the left panel. Similar phenomena are observed from theoretical studies in relativistic heavy-ion collisions (see, e.g., Ref.~\cite{Sun17}), while the spin polarization at large rapidities is so far still difficult to measure in relativistic heavy-ion collision experiments~\cite{STAR18}. Considering the uncertainty range of the spin-orbit coupling constant $W_0=80 \sim 150$ MeVfm$^5$~\cite{Les07,Zal08,Ben09}, a stronger spin-orbit coupling generally leads to a stronger global spin polarization in the $+y$ direction at midrapidites. Compared to the scenario with spin unchanged after collisions, the spin polarization for the spin randomized scenario is seen to be weaker at midrapidities due to the weaker spin polarization of the participant matter, and becomes negative at large rapidities due to the dominating effect from the negatively polarized spectator matter.

\begin{figure*}[ht]
	\includegraphics[scale=0.3]{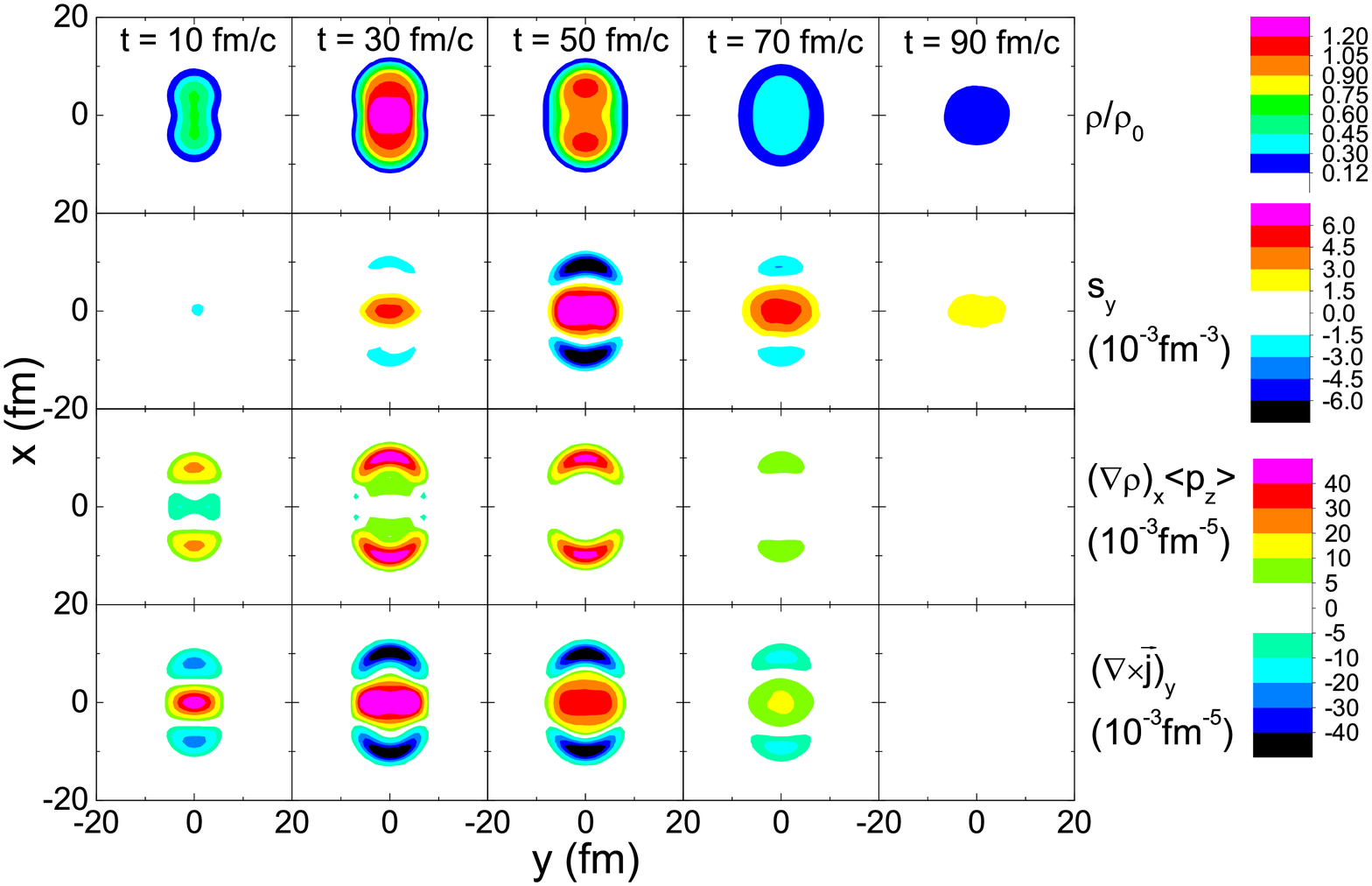}\includegraphics[scale=0.3]{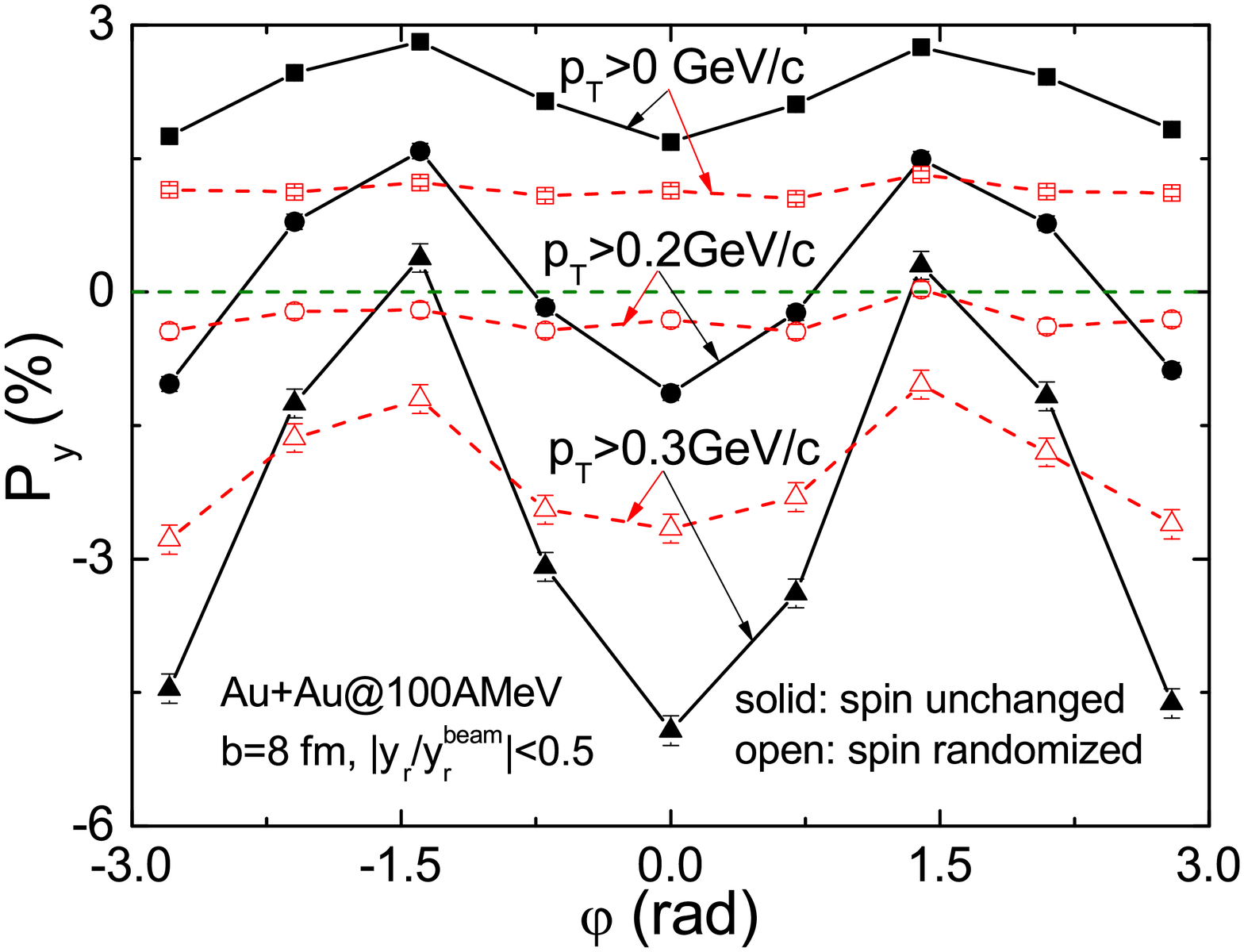}
	\caption{(Color online) Left: Density contours of the reduced nucleon number density (first row), $y$ component of the nucleon spin density (second row), $y$ component of the time-even spin-orbit potential (third row), and $y$ component of the time-odd spin-orbit potential in the transverse plane, for the spin unchanged scenario; Right: Spin polarization in $y$ direction of mid-rapidity free nucleons in different transverse momentum ranges and with different collision treatments as a function of the azimuthal angle. Results are for Au+Au collisions at the beam energy of 100 AMeV and an impact parameter $b=8$ fm.} \label{xoy_py}
\end{figure*}

The left panel of Fig.~\ref{xoy_py} shows similar quantities for the spin unchanged scenario but in the transverse plane, the time evolution of which are consistent with the section around $z\sim0$ in the corresponding plots in the left panel of Fig.~\ref{xoz}. Again, these density contours are similar for the scenario with the nucleon spin randomized after collisions, except that the spin polarization of the participant matter is weaker. The right panel of Fig.~\ref{xoy_py} displays the spin polarization of mid-rapidity free nucleons in different transverse momentum ($p_T$) ranges as a function of the azimuthal angle $\varphi=\text{atan2}(p_y,p_x)$. $\varphi \sim 0$ or $\pm \pi$ corresponds to nucleons emitted in the $\pm x$ direction, while $\varphi \sim \pm \pi/2$ corresponds to nucleons emitted in the $\pm y$ direction. As shown in the left panel of Fig.~\ref{xoy_py}, nucleons emitted in the $\pm x$ direction can be largely blocked by spectators, especially for high-$p_T$ nucleons which are emitted at earlier stages, leading to valleys at $\varphi \sim 0$ and $\pm \pi$ in the polarization distribution. For nucleons which are emitted in the $\pm y$ direction, they are not blocked by the spectator matter so their spin polarizations are mostly preserved, leading to double peaks around $\varphi \sim \pm \pi/2$. In addition, the blocking effect of the spectator matter on high-$p_T$ nucleons also leads to their weaker or even negative spin polarizations in $y$ direction. If the nucleon spin is randomized after collisions, the magnitude of the spin polarization becomes much weaker, while its angular dependence is also partially washed out, especially for low-$p_T$ nucleons that suffer from more collisions.

\begin{figure*}[ht]
	\includegraphics[scale=0.3]{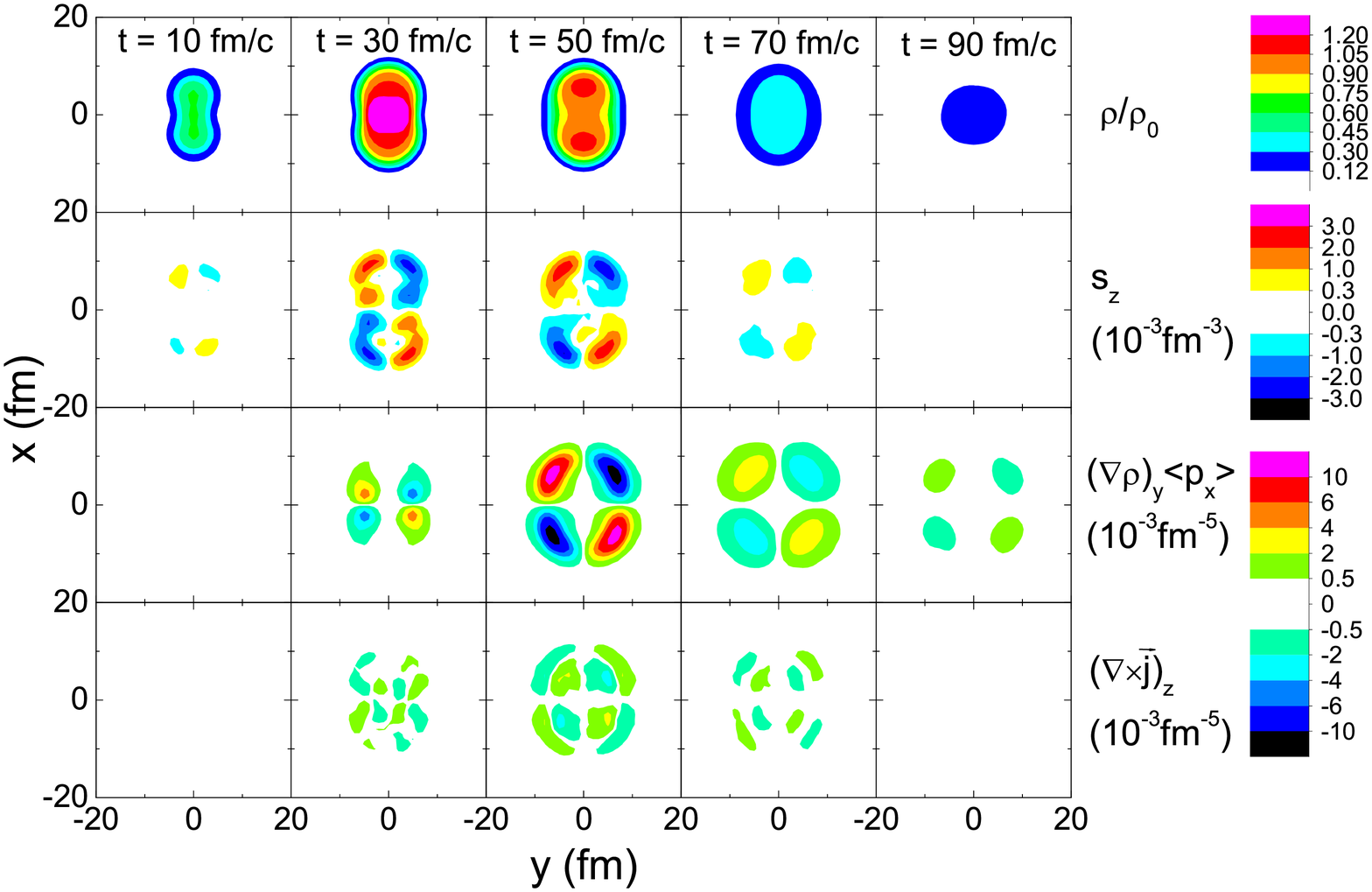}\includegraphics[scale=0.3]{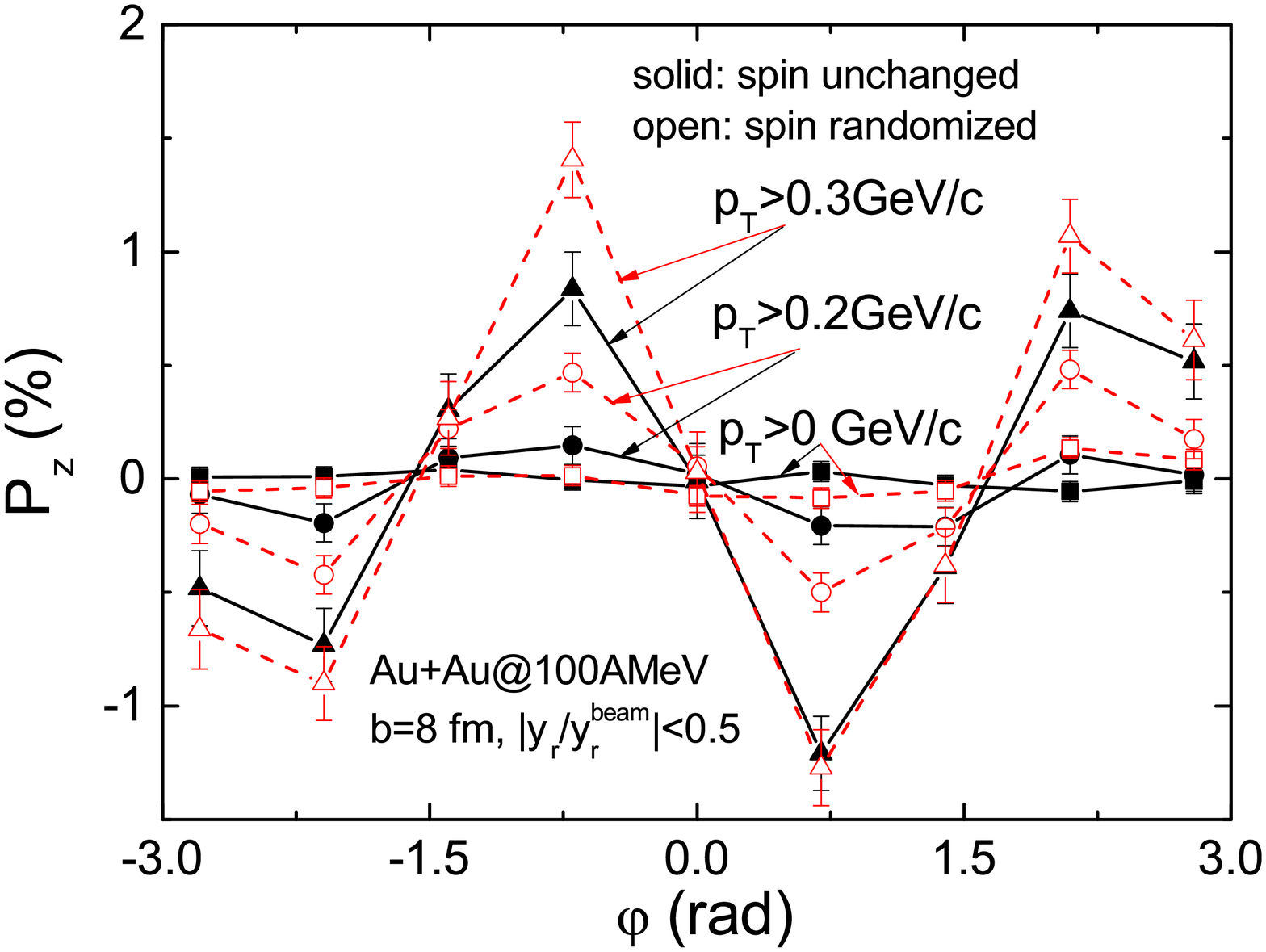}
	\caption{(Color online) Left: Density contours of the reduced nucleon number density (first row), $z$ component of the nucleon spin density (second row), $z$ component of the time-even spin-orbit potential (third row), and $z$ component of the time-odd spin-orbit potential in the transverse plane, for the spin unchanged scenario; Right: Spin polarization in $z$ direction of mid-rapidity free nucleons in different transverse momentum ranges and with different collision treatments as a function of the azimuthal angle. Results are for Au+Au collisions at the beam energy of 100 AMeV and an impact parameter $b=8$ fm.} \label{xoy_pz}
\end{figure*}

Figure~\ref{xoy_pz} displays the longitudinal spin polarization in the transverse plane as well as its azimuthal angular dependence. As shown in the second row of the left panel, the spin density in $z$ direction has some local structures, especially at the most compressed stage. Although the time-even component of the spin-orbit potential dominated by the $(\nabla \rho)_y \langle p_x\rangle$ term seems to have a stronger effect, as shown in the third row, it mostly affects the spectator region. After a close look at the contribution of the time-odd component of the spin-orbit potential, as shown in the fourth row, we find that it dominates the local spin polarization in the participant region. Compared to the corresponding contribution from the spin unchanged scenario shown in the left panel of Fig.~\ref{xoy_pz}, the spin randomized scenario leads to a stronger dissipation and a larger $(\nabla \times \vec{j})_z$ contribution in the participant region. The structure of the local spin polarization in $z$ direction observed in the left panel shows up in its azimuthal angular dependence as shown in the right panel. The polarization is positive in the ranges of $-\pi/2<\varphi<0$ and $\pi/2<\varphi<\pi$, and is negative in the ranges of $-\pi<\varphi<-\pi/2$ and $0<\varphi<\pi/2$, consistent with the spin polarization structure from the $(\nabla \times \vec{j})_z$ contribution in the participant region as shown in the left panel. A stronger local spin polarization is observed for high-$p_T$ free nucleons, which are emitted at earlier stages and thus carry the information of the spin density at that time. Different from the spin polarization in $y$ direction, the local spin polarization in $z$ direction seems to be enhanced if the nucleon spin is randomized after collisions. This is due to the larger $(\nabla \times \vec{j})_z$ contribution in the participant region for the spin randomized scenario. The azimuthal angular dependence of the local spin polarization in $z$ direction is similar to those from various theoretical approaches for relativistic heavy-ion collisions (see, e.g., Ref.~\cite{Xia18} and \cite{Bec18} for studies within the transport and hydrodynamic framework at $\sqrt{s_{NN}}=7.7-2760$ GeV), but has an opposite sign compared with the recent STAR measurement at $\sqrt{s_{NN}}=200$ GeV~\cite{STAR19}.

\begin{figure}[ht]
	\includegraphics[scale=0.3]{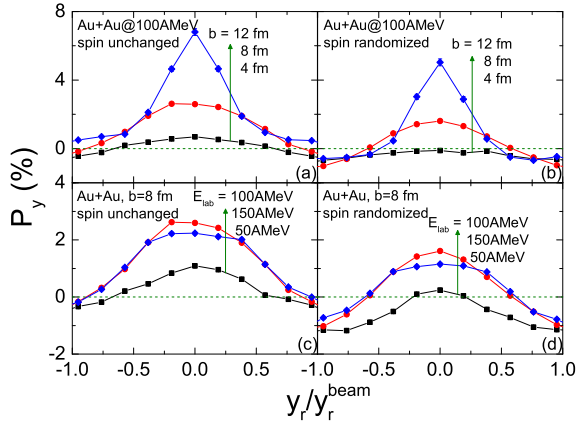}
	\caption{(Color online) Nucleon spin polarization in $y$ direction as a function of the reduced rapidity in Au+Au collisions at different beam energies and impact parameters with different treatments of the nucleon spin after nucleon-nucleon collisions.} \label{py_rap_bE}
\end{figure}

\begin{figure}[ht]
	\includegraphics[scale=0.3]{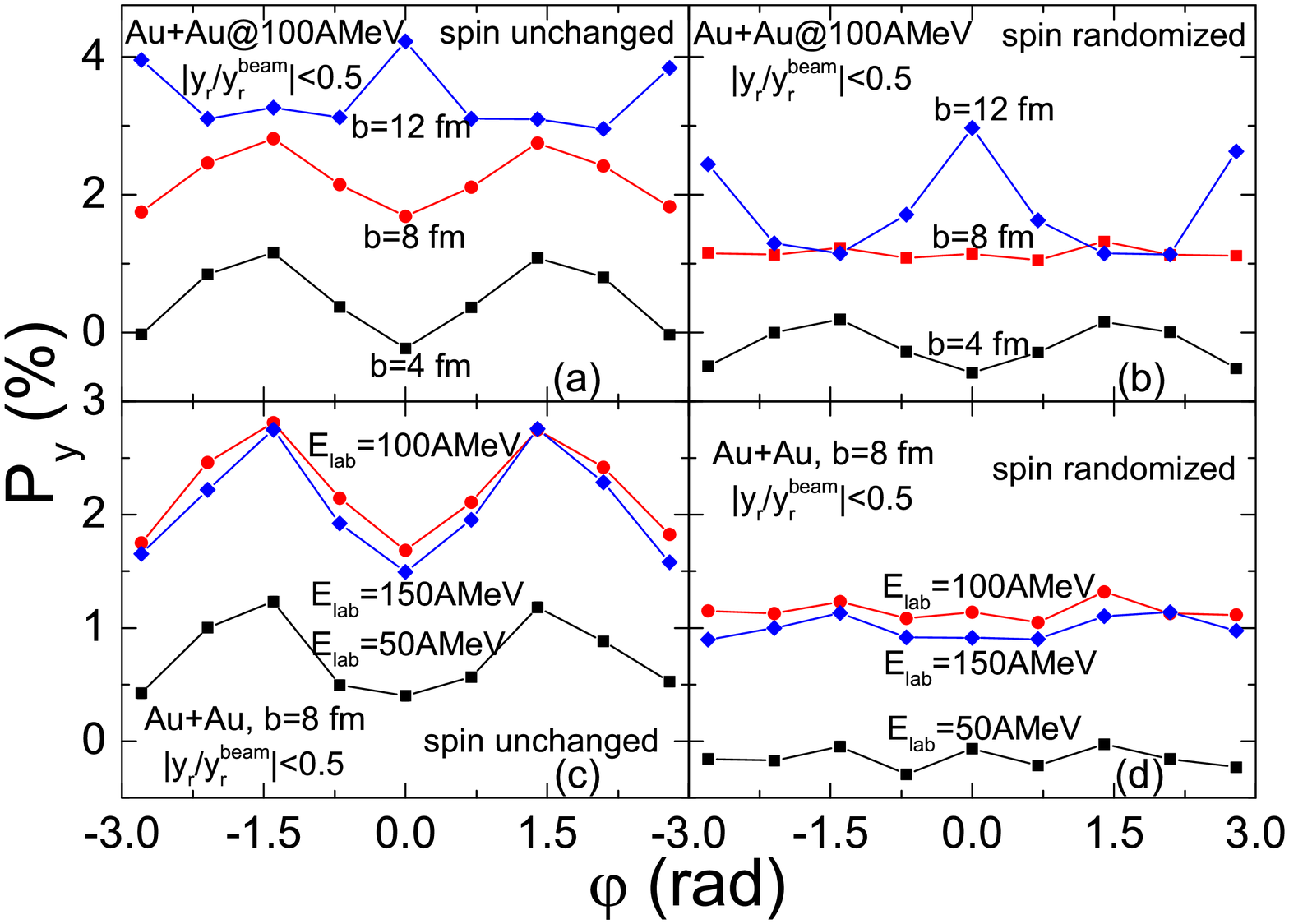}
	\caption{(Color online) Spin polarization in $y$ direction of mid-rapidity free nucleons as a function of the azimuthal angle in Au+Au collisions at different beam energies and impact parameters with different treatments of the nucleon spin after nucleon-nucleon collisions.} \label{py_phi_bE}
\end{figure}

\begin{figure}[ht]
	\includegraphics[scale=0.3]{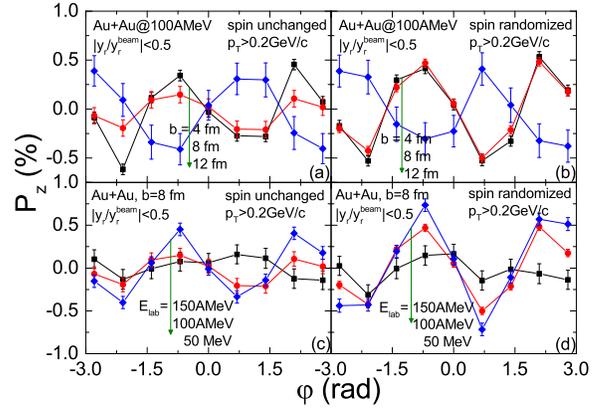}
	\caption{(Color online) Spin polarization in $z$ direction of mid-rapidity high-$p_T$ nucleons as a function of the azimuthal angle in Au+Au collisions at different beam energies and impact parameters with different treatments of the nucleon spin after nucleon-nucleon collisions.} \label{pz_phi_bE}
\end{figure}

So far we have illustrated the spin polarizations and discussed their mechanisms in Au+Au collision system at a fixed collision energy and for a given impact parameter, and we will further discuss the dependence of the spin polarization on the collision energy and the impact parameter. Figures~\ref{py_rap_bE} shows the global spin polarization of free nucleons in $y$ direction as a function of the reduced rapidity at different collision energies and impact parameters. Since the spin-orbit potential depends on the density gradient, the stronger spin polarization especially at midrapidities in peripheral collisions is understandable. Although the angular momentum is expected to be larger at higher collision energies, the effect of the spin-orbit potential saturates around the beam energy of $100\sim150$ AMeV, even with the spin unchanged scenario since the collisions thermalize the system anyway. These observations are consistent with our previous study on the spin-dependent collective flows~\cite{Xia14}, and the effect of the collision treatment is qualitatively similar in all cases. Figure~\ref{py_phi_bE} displays the azimuthal angular dependence of the spin polarization of mid-rapidity free nucleons in $y$ direction at different collision energies and impact parameters. It is interesting to see that the spin polarization at $b=12$ fm is not only stronger but also has a different azimuthal angular dependence, due to the different blocking effect from the spectator as well as the different shape of the polarized participant matter in peripheral collisions. Again, the magnitude and the azimuthal angular dependence are both suppressed if the nucleon spin is randomized after collisions. The local spin polarizations of mid-rapidity free nucleons in $z$ direction at different collision energies and impact parameters are displayed in Fig.~\ref{pz_phi_bE}. Interestingly, the local spin polarization at $b=12$ fm has an opposite sign in its azimuthal angular dependence compared with those at $b=4$ and 8 fm. This is again due to the different geometries in peripheral collisions, consistent with that observed in Fig.~\ref{py_phi_bE}. The local spin polarization is expected to saturate at even higher collision energies, and is stronger for the scenario with the nucleon spin randomized after collisions as a result of the larger $(\nabla \times \vec{j})_z$ contribution.

To summarize, based on a spin-dependent Boltzmann-Uehling-Uhlenbeck transport model, with the spin-dependent potential calculated from the lattice Hamiltonian method, and the spin-dependent differential cross sections and Pauli blockings incorporated, we have studied the nucleon spin polarization in non-central intermediate-energy heavy-ion collisions. Both the global spin polarization perpendicular to the reaction plane and the local spin polarization in the longitudinal direction are found to be mostly dominated by the time-odd component of the nuclear spin-orbit potential. The global spin polarization is seen to be stronger at midrapidities than at large rapidities, and shows some azimuthal angular dependence which can be understood from the blocking of the spectator matter. It is larger in peripheral collisions than in midcentral collisions, and saturates at the beam energy of around 100 AMeV. The local spin polarization is seen to be stronger at higher collision energies, and has an opposite sign in its azimuthal angular dependence in midcentral collisions compared with that observed in relativistic heavy-ion collisions, but this is not the case in peripheral collisions. The quantitative behavior of the spin polarization can also be affected by the strength of the nuclear spin-orbit interaction as well as the uncertainties of the in-medium nucleon spin flip after collisions. Our study helps to understand the spin polarization mechanisms in heavy-ion collisions dominated by nucleon degree of freedom. Our approach can be further extended to incorporate other hadrons together with their spin-orbit potentials, and be used to study the hadronic afterburner effect on spin polarizations in relativistic heavy-ion collisions, as well as spin polarization phenomena in heavy-ion collisions at the energy of a few GeV dominated by hadronic degrees of freedom, where hyperons or vector mesons can be produced and their spin polarizations can be experimentally measurable.

This work was supported by the National Natural Science Foundation of China under Grant No. 11922514.


\end{document}